\documentclass[sigconf]{acmart}

\usepackage{multirow}
\usepackage{amsmath}
\usepackage{enumitem}
\usepackage{array}

\renewcommand\footnotetextcopyrightpermission[1]{}  
\setcopyright{none}                       
\acmConference{}{}{}                      
\acmDOI{}
\acmISBN{}
\acmPrice{}

\author{Jie JW Wu}
\affiliation{%
  \institution{Michigan Technological University}
  \city{Houghton}
  \state{MI}
  \country{USA}
}
\email{jie.jw.wu@mtu.edu}

\author{Feiyu E}
\affiliation{%
  \institution{Michigan Technological University}
  \city{Houghton}
  \state{MI}
  \country{USA}
}
\email{efeiyu@mtu.edu}

\author{Bo Chen}
\affiliation{%
  \institution{Michigan Technological University}
  \city{Houghton}
  \state{MI}
  \country{USA}
}
\email{bchen@mtu.edu}

\title{Metamorphic Testing for Clinical ML Models:\\
  A Framework Proposal and Pilot Study}


\begin{document}

\begin{abstract}
Machine learning models for clinical prediction tasks such as in-hospital
mortality and sepsis onset routinely achieve high AUROC scores, yet AUROC
measures ranking correctness, not clinical sensibility. A model can rank
patients correctly in aggregate while predicting lower mortality risk when a
patient's SOFA score worsens, which contradicts established medical
guidelines. This paper proposes applying metamorphic testing (MT) to clinical
ML models as a way to check behavioral correctness without requiring
ground-truth labels for individual predictions. We design a catalog of 12
candidate metamorphic relations (MRs) for three ICU prediction tasks on
MIMIC-III/IV, each grounded in an authoritative clinical guideline. We also
propose a five-layer validation strategy for ensuring that MRs are clinically
sound before use. As a feasibility check, we run a pilot study on the UCI
Heart Disease dataset, where all three clinical models tested (AUROC
0.849--0.900) produce violation rates of 27--87\% on the five pilot MRs. An
injected-fault experiment shows that a sign-negation error in a blood pressure
feature went undetected by AUROC but produced a 31--67 percentage-point shift
in MT violation rate. These results suggest MT is a useful complement to
standard metrics for checking clinical model behavior.
\end{abstract}

\maketitle

\thispagestyle{plain}
\pagestyle{plain}

\renewcommand{\thefootnote}{}%
\footnotetext{Accepted at the AIware 2026 arXiv Track.}%
\renewcommand{\thefootnote}{\arabic{footnote}}

\section{Introduction}
\label{sec:intro}

The standard way to evaluate a clinical ML model is to measure its AUROC on
a held-out test set. This is useful, but it has an important limitation: AUROC
only captures whether the model ranks patients in the right order. It does not
check whether individual predictions are consistent with medical knowledge.

Consider a mortality prediction model with AUROC\,=\,0.87. Such a model could
still predict that a patient's risk decreases as their lactate rises into the
septic shock range, a behavior that contradicts clinical guidelines but leaves
AUROC unchanged. This limitation is an instance of the \emph{oracle
problem}~\cite{weyuker1982}: for most clinical inputs, we cannot verify an
individual prediction against ground truth.

Metamorphic testing (MT)~\cite{chen1998} addresses this by checking relational
properties across pairs of inputs rather than verifying individual outputs.
Given a semantically meaningful change to a patient record, an MT oracle
specifies how the model output should change. For example: if age increases
(all else constant), predicted mortality risk should not decrease. These
\emph{metamorphic relations} (MRs) encode clinical domain knowledge as
testable constraints and do not require labeled ground truth.

MT has been applied to traditional software~\cite{segura2016}, scientific
computing~\cite{kanewala2014}, and more recently to healthcare ML on medical
imaging~\cite{mtbreast2020} and clinical NLP~\cite{jaganathan2025}. Its
application to structured EHR time-series prediction, the dominant data format
in ICU clinical AI, has not been explored.

This paper makes the following contributions. First, we propose a framework,
\textsc{ClinMT}, that applies MT to clinical ML models trained on MIMIC-III/IV
structured data (\S\ref{sec:framework}). Second, we design a catalog of 12
candidate MRs covering three ICU tasks (mortality, decompensation, sepsis
onset), each traceable to a clinical guideline (\S\ref{sec:mrs}). Third, we
report a pilot study on the UCI Heart Disease dataset that checks whether MR
violations can be observed in practice, and whether they reveal model
behaviors that AUROC misses (\S\ref{sec:experiments}). The pilot serves as a
controlled first step before scaling to the more complex MIMIC-III setting.

\section{Background}
\label{sec:background}

\subsection{Metamorphic Testing}

Given a program $f$ whose outputs cannot be individually verified, metamorphic
testing~\cite{chen1998} checks whether pairs of related inputs produce outputs
that satisfy a known relational constraint, called a \emph{metamorphic
relation} (MR). For a clinical ML model $f$ and patient record $x$, a
monotone increasing MR takes the form:
\begin{equation}
  \phi(x) \;\Rightarrow\; f(\phi(x)) \geq f(x) - \epsilon
\end{equation}
where $\phi$ is a perturbation (e.g., increasing age by 10 years) and
$\epsilon$ is a tolerance. An \emph{MR violation} occurs when the output
relation fails to hold. Segura et al.~\cite{segura2016} survey MT across
domains and show that it reliably detects faults in the absence of a
conventional oracle.

\subsection{Clinical ML on MIMIC}

MIMIC-III and MIMIC-IV~\cite{johnson2016mimic,johnson2023mimic} are
de-identified ICU databases containing over 40,000 admissions with hourly
time-series of vitals, labs, and clinical interventions. Harutyunyan et
al.~\cite{harutyunyan2019} define four benchmark prediction tasks on MIMIC-III
(in-hospital mortality, decompensation, length of stay, and phenotyping) and
provide standardized preprocessing and LSTM baselines. Our MR catalog targets
the first three tasks, since they have direct clinical relevance and clear
outcome semantics.

\section{The \textsc{ClinMT} Framework}
\label{sec:framework}

\begin{figure}[ht]
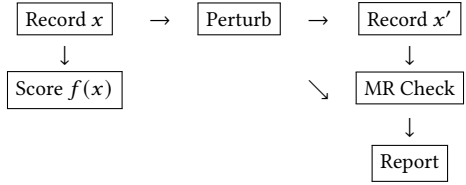

\centering
\small
\renewcommand{\arraystretch}{1.3}
\begin{tabular}{ccccc}
\fbox{Record $x$}
  & $\to$
  & \fbox{Perturb}
  & $\to$
  & \fbox{Record $x'$} \\
$\downarrow$ & & & & $\downarrow$ \\
\fbox{Score $f(x)$} & & & $\searrow$ & \fbox{MR Check} \\
& & & & $\downarrow$ \\
& & & & \fbox{Report} \\
\end{tabular}
\caption{\textsc{ClinMT} pipeline: both original and perturbed records are
  scored; the MR checker records violations.}
\Description{Flow diagram: Record x feeds into a Perturb step to produce
  Record x-prime. Both are scored by the model. The two scores are passed to
  an MR Check step, which outputs a Report.}
\label{fig:pipeline}
\end{figure}

Figure~\ref{fig:pipeline} shows the \textsc{ClinMT} testing pipeline. Given a
trained model and a test set, the framework iterates over each patient record
and each MR. A perturbation engine produces a modified record; both records
are scored; the MR checker applies the relation and records whether it holds.

\subsection{Five-Layer MR Validation Strategy}

Not every clinically plausible perturbation produces a reliable MR. We propose
five validation steps before adding an MR to the catalog. Steps L1, L4, and
L5 are applied in the current work; L2 and L3 are planned for the larger-scale
MIMIC experiment and require clinician involvement.

\textbf{L1 -- Clinical guideline grounding.} Each MR cites an authoritative
source such as Sepsis-3~\cite{singer2016sepsis3}, the Surviving Sepsis
Campaign~\cite{evans2021ssc}, or APACHE\,II~\cite{knaus1985apache}, and
includes a plain-language rationale.

\textbf{L2 -- Clinician review (planned).} Two or three ICU physicians review
each candidate MR through a structured survey. MRs with less than 80\%
agreement are revised or removed. Cohen's $\kappa$ will be reported.

\textbf{L3 -- Empirical direction check (planned).} We will verify the
claimed direction of each MR using Spearman correlation and Wilcoxon
signed-rank tests on a held-out MIMIC-III cohort, to confirm the relation
holds in real clinical data before using it as a test oracle.

\textbf{L4 -- Perturbation bounds.} Perturbation magnitudes are chosen based
on clinical conventions, not sampled randomly. A plausibility filter then
excludes any record whose perturbed feature value falls outside absolute
physiological limits, ensuring the perturbed input represents a feasible
patient state.

\textbf{L5 -- Formal type assignment.} Each MR is assigned one of three
formal types:
\begin{itemize}[noitemsep, leftmargin=*, topsep=2pt]
  \item \textbf{Monotone Increasing (MI):} $f(x') \geq f(x) - \epsilon$ when
        feature $k$ increases.
  \item \textbf{Monotone Decreasing (MD):} $f(x') \leq f(x) + \epsilon$ when
        $k$ increases.
  \item \textbf{Invariance (INV):} $|f(x') - f(x)| < \epsilon$ under a
        perturbation that should not affect the prediction.
\end{itemize}
We use a fixed tolerance of $\epsilon = 0.01$ for monotone MRs (MI/MD) and
$\epsilon = 0.05$ for invariance MRs (INV).

\section{Metamorphic Relation Catalog}
\label{sec:mrs}

Table~\ref{tab:mrs} lists the 12 candidate MRs we have designed for
MIMIC-III/IV tasks. Each MR has passed L1 (guideline grounding); L2 clinician
review and L3 empirical checks are planned for the MIMIC-III experiment.

\begin{table}[ht]
\centering
\caption{Proposed MR catalog for MIMIC-III/IV. L1 grounding complete;
  L2/L3 planned for MIMIC-III experiment. MI\,=\,Monotone Increasing; INV\,=\,Invariance.}
\label{tab:mrs}
\small
\setlength{\tabcolsep}{3pt}
\begin{tabular}{@{}llp{2.8cm}p{1.3cm}@{}}
\toprule
\textbf{ID} & \textbf{Type} & \textbf{Perturbation} & \textbf{Source} \\
\midrule
\multicolumn{4}{@{}l}{\textit{In-Hospital Mortality}} \\
MR-M1 & MI  & Age $\uparrow$ (+10\,yr)            & \cite{knaus1985apache}   \\
MR-M2 & MI  & SOFA $\uparrow$ (+2 pts)             & \cite{singer2016sepsis3} \\
MR-M3 & MI  & Lactate $\uparrow$ ($>$4\,mmol/L)   & \cite{evans2021ssc}      \\
MR-M4 & INV & ICU unit type swapped                & Consensus                \\
\midrule
\multicolumn{4}{@{}l}{\textit{Decompensation}} \\
MR-D1 & MI  & SpO$_2$ $\downarrow$ ($-5\%$)        & \cite{aarc2002}          \\
MR-D2 & MI  & Resp.\ rate $\uparrow$ ($>$25/min)  & \cite{evans2021ssc}      \\
MR-D3 & MI  & GCS $\downarrow$ ($-3$ pts)          & \cite{knaus1985apache}   \\
MR-D4 & MI  & MAP $\downarrow$ ($<$65\,mmHg)       & \cite{evans2021ssc}      \\
\midrule
\multicolumn{4}{@{}l}{\textit{Sepsis Onset}} \\
MR-S1 & MI  & HR $\uparrow$ + SBP $\downarrow$ together  & \cite{singer2016sepsis3} \\
MR-S2 & MI  & Temp.\ $>38.3$ or $<36\,^{\circ}$C  & \cite{singer2016sepsis3} \\
MR-S3 & MI  & Infection + organ dysfunction         & \cite{singer2016sepsis3} \\
MR-S4 & INV & Race/ethnicity field changed          & \cite{chen2021disparities} \\
\bottomrule
\end{tabular}
\end{table}

\noindent\textbf{Fairness as invariance.} MR-S4 and MR-M4 are invariance MRs
over demographic and administrative attributes. Demographic fairness testing
can be viewed as a special case of MT: an invariance MR restricted to
protected attributes. This framing lets \textsc{ClinMT} subsume fairness
checks within the same testing workflow~\cite{roosli2022,vanschaik2024}.

\noindent\textbf{Fault types.} Table~\ref{tab:bugs} lists six categories of
model faults that MR violations can surface. Several are not detectable by
AUROC alone, because they affect individual-level prediction directions rather
than population-level ranking order.

\begin{table}[ht]
\centering
\caption{Fault types that MR violations can detect.}
\label{tab:bugs}
\small
\setlength{\tabcolsep}{3pt}
\begin{tabular}{@{}lp{4.0cm}@{}}
\toprule
\textbf{Fault type} & \textbf{Example} \\
\midrule
Spurious correlation & Age--mortality inversion in a trauma subgroup \\
Feature interaction  & High lactate and low MAP together underweighted \\
Temporal ordering    & Declining SpO$_2$ not distinguished from rising \\
Preprocessing error  & Fahrenheit/Celsius mix in temperature field \\
Calibration fault$^*$& High- and low-risk patients both score below 0.10 \\
Sign / index error   & SOFA coefficient accidentally negated \\
\bottomrule
\multicolumn{2}{@{}l}{\small $^*$ Ranking correct, magnitudes wrong; AUROC does not detect.}
\end{tabular}
\end{table}

\section{Pilot Study}
\label{sec:experiments}

\subsection{Setup}

We report a pilot study to check whether the MR framework produces observable
violations on real clinical data and whether MT is sensitive to an injected
model fault that AUROC misses.

\textbf{Dataset.} We use the UCI Heart Disease (Cleveland) dataset~\cite{uci_heart}:
303 patients, 13 input features, binary outcome (disease present or absent), 54.5\% positive rate. This is a simpler dataset than our intended
target (MIMIC ICU time-series), but it is freely available without a data use
agreement and contains several features with clear monotone or invariance
properties under clinical reasoning. We use it as a deliberate pilot step
before conducting the larger-scale experiment on MIMIC ICU data.

\textbf{Pilot MRs.} We define five MRs for the cardiovascular domain, each of
which maps to a MIMIC catalog MR type:
MR-H1 (age$\uparrow$ +5\,yr, MI, maps to MR-M1),
MR-H2 (resting BP$\uparrow$ +15\,mmHg, MI, maps to MR-D4),
MR-H3 (cholesterol$\uparrow$ +30\,mg/dl, MI),
MR-H4 (ST depression$\uparrow$ +1.0, MI, maps to MR-M2), and
MR-H5 (sex swapped, INV, maps to MR-S4).
Perturbation magnitudes are fixed increments chosen to represent clinically
meaningful changes: 5\,years corresponds to a standard age-risk interval,
15\,mmHg to a one-stage blood pressure elevation, 30\,mg/dl to a moderate
cholesterol increase, and 1.0 to a clinically notable ST-depression shift.
The L4 plausibility filter uses absolute physiological bounds (age 18--100,
trestbps 80--220\,mmHg, chol 100--600\,mg/dl, oldpeak 0--8) to exclude any
patient record where the perturbed value would be physiologically implausible;
all 61 test patients passed this filter for every MR.

\textbf{Models.} We train three standard classifiers with an 80/20 stratified
split: logistic regression (LR) with balanced class weights, random forest
(RF) with 100 trees, and a two-layer MLP (32, 16 hidden units; early
stopping). All models include feature standardization (zero mean, unit
variance) as a preprocessing step. The test set has 61 patients.

\textbf{Injected-fault protocol.} To evaluate MT's sensitivity to a
preprocessing error, we simulate a sign-negation fault: we negate the resting
BP feature in both the training and test data, retrain the model on the
corrupted training data, and evaluate it on the correspondingly corrupted test
data. We then compare $|\Delta\mathrm{AUROC}|$ and $|\Delta\mathrm{VR}|$ on
MR-H2 between the original and faulty model variants.

\subsection{Results}

\textbf{Violation rates.} Table~\ref{tab:results} shows MR violation rates for
all three models. Despite AUROC values between 0.849 and 0.900, all three models
produce non-zero violation rates on four of five MRs. The highest rates appear
on MR-H4 (ST depression increase should not lower risk), where all models
violate the relation in 77--87\% of applicable test cases. MR-H5 (sex should
not change the prediction) shows rates of 28--87\%. These figures indicate
that, even on a simple dataset, models with competitive AUROC can behave
inconsistently with clinical expectations across a substantial fraction of
test cases.

The logistic regression model's 0\% violation rate on MR-H1 (age) reflects
that LR happens to learn a positive coefficient for age on this split, so the
monotone relation holds by construction. The high violation rates on MR-H3 and
MR-H4 for LR are due to negative learned coefficients for cholesterol and ST
depression, indicating the model has learned directions that conflict with
clinical knowledge despite high ranking accuracy.

\begin{table}[ht]
\centering
\caption{MR violation rates (\%) on UCI Heart Disease test set (61 patients).
  AUROC values shown below model names.}
\label{tab:results}
\small
\setlength{\tabcolsep}{4pt}
\begin{tabular}{@{}lp{2.4cm}ccc@{}}
\toprule
\textbf{MR} & \textbf{Perturbation} & \textbf{LR} & \textbf{RF} & \textbf{MLP} \\
            & \textit{(AUROC)}      & \small 0.871 & \small 0.900 & \small 0.849 \\
\midrule
MR-H1 & Age $\uparrow$ (+5\,yr)        &  0.0\% & 32.8\% & 39.3\% \\
MR-H2 & Resting BP $\uparrow$ (+15)    & 67.2\% & 31.1\% & 49.2\% \\
MR-H3 & Cholesterol $\uparrow$ (+30)   & 80.3\% & 45.9\% & 29.5\% \\
MR-H4 & ST depression $\uparrow$ (+1)  & \textbf{86.9\%} & \textbf{77.0\%} & \textbf{77.0\%} \\
MR-H5 & Sex swapped (INV)              & 86.9\% & 63.9\% & 27.9\% \\
\bottomrule
\end{tabular}
\end{table}

\textbf{Injected-fault results.} Table~\ref{tab:fault} shows that the
sign-negation fault has a small effect on AUROC: at most 0.007 for MLP and
zero for LR. The MT framework, however, produces a qualitatively different
signal. For the faulty model, the sign-negated resting BP values are negative
(e.g., $-130$\,mmHg); when MR-H2 then increments these by $+15$, the perturbed
values ($-115$\,mmHg) fall below the physiological lower bound of 80\,mmHg, so
the L4 plausibility filter excludes every test case ($n_\mathrm{tested} = 0$,
shown as VR$_F$\,=\,0 in Table~\ref{tab:fault}). This collapse in testable
cases is itself a detectable signal: the fault renders the affected MR
unevaluable. On the healthy model, the same filter passes all 61 cases and
finds violation rates of 31--67\% on MR-H2.

We report the sensitivity ratio $S =
|\Delta\mathrm{VR}_\mathrm{pp}|/(|\Delta\mathrm{AUC}| \times 100)$ as a coarse
measure of relative sensitivity. For RF and MLP, $S$ is 141 and 76
respectively. For LR, $|\Delta\mathrm{AUC}| = 0$, so $S$ is undefined (AUROC
provides no signal at all); the release script guards this division with a
small constant rather than leaving it undefined, so the raw output file
reports a large finite placeholder for this cell instead of the
\texttt{undefined} we report here. These numbers should be interpreted with caution
given the small test set size and single fault type; they are indicative
rather than conclusive.

\begin{table}[ht]
\centering
\caption{Injected-fault experiment: resting BP sign negation (model retrained
  on corrupted data). AUROC changes are small ($\leq$0.007); VR$_F$\,=\,0 because
  the L4 plausibility filter excludes all faulty-model test cases (see footnote).}
\label{tab:fault}
\footnotesize
\setlength{\tabcolsep}{2pt}
\begin{tabular}{@{}lccccccc@{}}
\toprule
\textbf{Model} & AUC$_H$ & AUC$_F$ & $|\Delta\mathrm{AUC}|$ & VR$_H$ & VR$_F$ & $|\Delta\mathrm{VR}|$ & $S$ \\
\midrule
LR  & 0.871 & 0.871 & 0.000 & 67.2\% & 0\%$^\dagger$ & 67.2\,pp & undef. \\
RF  & 0.900 & 0.903 & 0.002 & 31.1\% & 0\%$^\dagger$ & 31.1\,pp & 141$\times$ \\
MLP & 0.849 & 0.842 & 0.007 & 49.2\% & 0\%$^\dagger$ & 49.2\,pp &  76$\times$ \\
\bottomrule
\multicolumn{8}{@{}p{0.95\columnwidth}}{\small $H$\,=\,original model;
  $F$\,=\,faulty model (retrained on negated BP);
  $S = |\Delta\mathrm{VR}|_\mathrm{pp}/(|\Delta\mathrm{AUC}|\times 100)$.
  $^\dagger$Sign negation produces negative BP values ($<0$\,mmHg);
  adding $+15$ still leaves them below the physiological bound of 80\,mmHg,
  so the L4 filter excludes all cases ($n_\mathrm{tested}$\,=\,0).
  VR$_F$\,=\,0 does not mean the faulty model passes MR-H2; the MR
  cannot be evaluated on physiologically impossible inputs.}
\end{tabular}
\end{table}

\section{Related Work}
\label{sec:related}

\textbf{MT in healthcare ML.} Goens et al.~\cite{mtbreast2020} apply MT to
breast cancer image classifiers and Rehman and Izurieta~\cite{mtpneumonia2025}
apply a statistical MT approach to CNN-based image classifiers. Jaganathan
et al.~\cite{jaganathan2025} apply MT to clinical NLP for automated ICD
coding by defining text-level perturbations.
These works target image and text modalities. Our work targets structured EHR
time-series, which require domain-specific feature-level MRs grounded in
quantitative clinical thresholds.

\textbf{Testing MIMIC-based models.} Prior work on MIMIC model evaluation
focuses on subgroup performance and demographic
fairness~\cite{roosli2022,vanschaik2024}. \textsc{ClinMT} complements this by
checking behavioral consistency across the clinical feature space, not just
across demographic groups.

\textbf{ML testing more broadly.} Zhang et al.~\cite{zhang2020mltest} survey
testing techniques for ML systems including MT and metamorphic
oracles~\cite{dwarakanath2018}. Our contribution is not a new testing
technique but a domain-specific instantiation for clinical EHR models, with
MRs tied to clinical guidelines and a protocol for validating them before use.
The hidden-debt framing of Sculley et al.~\cite{sculley2015} and the
dataset-shift work of Finlayson et al.~\cite{finlayson2021} motivate why
behavioral testing is needed alongside standard evaluation.

\section{Threats to Validity}
\label{sec:threats}

\textbf{External validity.} The pilot study uses the UCI Heart Disease dataset
(303 patients, 61 test cases), which is simpler than the intended MIMIC ICU
setting in scale, feature complexity, and task structure. We cannot claim that
the violation rates observed here will transfer to MIMIC models, and the
sensitivity ratio numbers are derived from a single injected fault on a small
test set.

\textbf{Construct validity.} MR-H2 holds that increasing resting BP should
not lower heart disease risk. This is supported by epidemiological evidence,
but the relation may not hold uniformly across all patient subgroups or
clinical contexts. Layers L2 and L3 of our validation strategy are designed
to surface such exceptions, but they have not yet been applied to the pilot
MRs.

\textbf{Internal validity.} The injected fault (sign negation of resting BP
in training and test data) is a synthetic scenario. Real-world preprocessing
errors may behave differently.

\textbf{Conclusion validity.} The 76--141$\times$ sensitivity ratio is an
informal comparison between two metrics with different scales. It is reported
to illustrate the qualitative difference in sensitivity; it should not be
interpreted as a precise quantitative claim.

\section{Conclusion}
\label{sec:conclusion}

This paper proposes applying metamorphic testing to clinical ML models trained
on structured EHR data. We designed a catalog of 12 candidate MRs for three
MIMIC-III/IV prediction tasks, each grounded in a clinical guideline. A pilot
study on UCI Heart Disease shows that models with AUROC between 0.849 and 0.900
can still violate medically grounded behavioral constraints at high rates, and
that a sign-negation preprocessing fault that AUROC cannot detect produces a
large shift in MT violation rate. These are preliminary results on a simpler
dataset; the primary purpose of the pilot is to confirm that the MR-based
testing approach is implementable and produces non-trivial signal.

Next steps include applying the framework to MIMIC-III benchmark LSTM models
from Harutyunyan et al.~\cite{harutyunyan2019},
conducting clinician review of the proposed MR catalog (L2), and running the
empirical direction check (L3) on held-out MIMIC data.

\bibliographystyle{ACM-Reference-Format}
\bibliography{references}

\end{document}